\documentclass[apjl]{emulateapj}

\shorttitle{LITHIUM ABUNDANCES IN RED GIANTS OF M4}
\shortauthors{D'ORAZI \& MARINO}

\begin{document}


\title{LITHIUM ABUNDANCES IN RED GIANTS OF M4:\\ EVIDENCE FOR ASYMPTOTIC GIANT BRANCH STAR POLLUTION IN GLOBULAR CLUSTERS?}


\author{Valentina D'Orazi\altaffilmark{1}}
\author{Anna F. Marino\altaffilmark{2}}

\altaffiltext{1}{INAF--Osservatorio Astronomico di Padova, 
vicolo dell'Osservatorio 5 , I-35122, Padova, Italy}
\altaffiltext{2}{Dipartimento di Astronomia, Universit\`a di Padova, vicolo
dell'Osservatorio 3, I-35122, Padova, Italy}
\email{valentina.dorazi@oapd.inaf.it, anna.marino@unipd.it}


\begin{abstract}
The determination of Li and proton-capture element abundances in globular cluster (GC) giants allows us to
constrain several key questions on the multiple population scenarios in GCs, from  
formation and early evolution, to pollution and dilution mechanisms. 
In this Letter we present our results on Li abundances for a large sample of giants in the
intermediate-metallicity GC NGC~6121 (M4), for which  Na and O have been already determined by Marino et
al.
The stars analyzed are both below and above the red giant branch bump luminosity. 
We found that the first and second generation stars share the same Li
content, suggesting that a Li production must have occurred. This is a strong observational evidence providing support for the scenario 
in which asymptotic giant branch stars are GC polluters.

\end{abstract}


\section{INTRODUCTION}\label{sec:intro}
The presence of multiple populations as a characterizing property of globular clusters (GCs)
is widely accepted nowadays. Since the first photometric and spectroscopic studies, 
which revealed multiple sequences (e.g. Dickens \& Woolley, 1967; Lee et al. 1999, 2009; Bedin et al. 2004) 
and/or light element (anti)correlations (see Gratton et al. 2004), it became evident
that GC stars are neither coeval nor chemically homogeneous. Hence, GCs host at least two stellar generations. 
Thanks to the advent of 8$-$10 m class telescopes, which also allowed 
targeting fainter main-sequence (MS) stars, several abundance studies have 
shown that chemical (anti)correlations are also present in unevolved (MS/turnoff (TO)) or scarcely
evolved (SGB) stars (Gratton et al. 2001; Carretta et al. 2004; Pasquini et al. 2005). 
This evidence implies that a previous generation of stars has
activated CNO, NeNa, and MgAl cycles in their interiors, in order to deplete O and Mg and 
enhance Na and Al,
respectively. The origin/nature of such kind of stars is still debated with   
two coexistent scenarios: (1) intermediate-mass asymptotic giant branch (AGB) stars in hot bottom burning phase (Ventura \&
D'Antona 2009) and (2) fast rotating massive stars (FRMS; Decressin et al. 2007).

Our group has recently carried out an extensive survey, focusing 
on the determination of
proton-capture elements in 19 GCs (Carretta et al. 2009c),
with the main objective of discovering and
understanding the nature (and the extent) of the chemical (anti)correlations and their
link with global cluster parameters (horizontal branch (HB) morphology, metallicity, etc.). 
The large sample of stars ($\sim$1200 GC members), analyzed in a very homogeneous 
and accurate way,
revealed that while the Na$-$O anticorrelation is present in all the GCs
(i.e. the second generation is not a ``perturbation"), the shape of the
Na$-$O distribution varies from cluster to cluster (Carretta et al. 2009c).
On the other hand, the analysis of UVES  spectra for $\sim$
200 stars by Carretta et al. (2009b) has shown that the Mg$-$Al anticorrelation is not present
in all GCs. Both these indications suggest 
that the typical polluter masses change from
cluster to cluster: this variation is apparently driven by a combination of
cluster luminosity and metallicity.

In this context, lithium abundances offer a complementary approach to $p$-capture
elements allowing to address several important issues.
If no Li is produced by the polluters, the multiple population scenario predicts that Li and O are positively correlated, 
while Li and Na
anticorrelated. Na-poor, O/Li-rich stars are the first population born in the cluster 
(they share the same chemical composition of field stars at the same metallicity), and 
Na-rich, O/Li-poor stars constitute the second generation. 
Within the same hypothesis, Li is an excellent tracer of the dilution process acting within each star: 
only through Li abundance determinations we can determine the amount of pristine (and of polluted)
material present in each star. In particular, we hope to answer two fundamental questions: 
do 100\% polluted stars (Li$\sim$0) exist or even the most extreme population 
still contains a certain fraction of primordial matter? Also, is the minimum measurable Li content
the same for all GCs or does it vary from cluster to cluster?

On the other hand, Li offers the exciting chance to observationally constrain the nature
of the polluters. If the progenitors of second generation stars are FRMS, they have destroyed their original Li content. 
On the other hand, if AGB stars are responsible for intra-cluster pollution, they may have non-negligible Li yield, given
the Li production via the ``$^7$Be transport" mechanism (Cameron \& Fowler 1971).
As a consequence, we can reveal if AGB stars are responsible for GC pollution
through two main observational evidences: (1) the presence of very Li-rich stars among GC populations and 
(2) the lack of Li$-$Na anti-correlation (or Li$-$O correlation), with the second generation stars also showing a rather high 
Li content. 
In a recent work, we focused on $\sim$90 TO stars belonging
to the metal-rich GC 47 Tuc: in this case, likely because of the high metallicity, a
large star-to-star scatter in Li abundances erases any Li$-$Na anticorrelation, while Li and O appear 
to be only weakly positively correlated (D'Orazi et al. 2010). 
The cluster seems to display a different behaviour from NGC~6752 and NGC~6397: for the first one, Pasquini et
al. (2005) detected a significant Li$-$Na anticorrelation (and also Li$-$O correlation and Li$-$N
anticorrelation). Concerning the second cluster, the situation seems more complex since Lind et al. (2009)
detected a quite constant value in Li abundances, with only three stars (out of 100) driving a hint of Li$-$Na anticorrelation.
Enlarging the sample of simultaneous determinations of Li and $p$-capture elements in GCs is hence of paramount importance:
in this Letter we present Li results on the intermediate-metallicity ([Fe/H]=-1.18; Carretta et al.\ 2009a) cluster NGC~6121 (M4), by
analyzing the same sample of a hundred red giant branch (RGB) stars studied by Marino et al.\ (2008).
As found by Marino and coworkers, M4 hosts two distinct populations of stars, 
mainly characterized by a different sodium content (i.e., the Na-rich and Na-poor groups), and defining different sequences 
on the color$-$magnitude diagram $U$ versus ($U-B$). 
We derived Li abundances for the stars belonging to the two groups, and located both below and above the RGB bump luminosity;
this evolutionary stage plays a fundamental role in this context. 

Theoretical models (Iben 1967), confirmed by observations of field stars by Gratton et al. (2000), predict 
a depletion in Li due to the first dredge-up (1DUP) of a factor 
of $\sim$ 20 at the base of the SGB branch.
On the lower RGB, below the bump luminosity, the molecular
weight gradient associated with the H abundance jump acts
as a barrier that prevents further extra mixing (Charbonnel et al. 1994); then 
the Li abundance remains constant until the
RGB bump is reached. At this stage, the H shell reaches and cancels this discontinuity and 
non-standard mixing processes (non-convective extramixing, see, e.g., Charbonnel \& Zahn 2007) 
cause the total destruction of
the remaining Li.

\section{SAMPLE AND ANALYSIS}\label{sec:analysis}
Our sample consists of 104 RGB stars, whose spectra were acquired with FLAMES on VLT/UT2 (Pasquini et al. 2002)
with the fiber link to the high-resolution spectrograph UVES ($R\sim40,000$). 
A detailed description of observations, target properties, and data reduction and derivation of atmospheric parameters is provided in 
Marino et al. (2008).
Adopting Kurucz (1993) model atmospheres and using the ROSA abundance code
(Gratton, 1988), we derived Li abundances by means of a spectral 
synthesis of the Li~{\sc i} resonance doublet at 6708 \AA.  
We changed the CN values for the two different groups of Na-rich and Na-poor stars (threshold value at [Na/Fe]=0.2 dex) 
in order to optimize the synthesis best fit and to account for the 
CN enhancement in the Na-rich population (see Marino et al. 2008). 

Abundances for Na{\sc i} along with stellar parameters and metallicity are the ones 
presented in Marino et al. (2008).
Concerning Li abundance, error estimates have been computed in the same fashion as described in D'Orazi et al. (2010), taking into account
both stellar parameter and best-fit uncertainties; for errors in 
Na (internal and systematic), we refer the reader to Marino et al. (2008). Stellar parameters and abundances
are given in Table~\ref{t:tab1} (completely available in electronic version through CDS).

\section{RESULTS AND DISCUSSION}\label{sec:disc}

\begin{center}
\begin{figure}
\includegraphics[width=9cm]{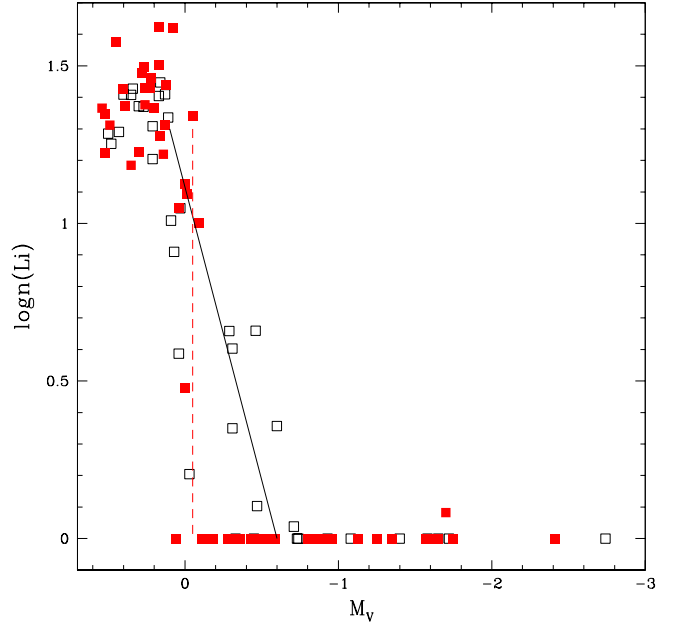}
\caption{Li as a function of absolute magnitudes for Na-rich (filled squares) and
Na-poor stars (empty squares). The dashed line marks the bump luminosity
as derived by Ferraro et al. (1999), while the solid line is an eye fit
to Na-poor population. (A color version of this figure is available in the online journal.)}\label{f:limag}
\end{figure}
\end{center}

In Figure~\ref{f:limag}, we show the Li abundances as a function of the absolute magnitude 
$M_{\rm V}$ for all our sample stars: as expected, Li disappears above the bump luminosity 
($M_{\rm V}$=$-$0.05$\pm$0.10; Ferraro et al. 1999). 
If we focus on the region below the bump level (at the left side of the dashed line in Figure~\ref{f:limag}), there is
no systematic difference in Li abundances between Na-rich (filled squares) and Na-poor stars (empty symbols).
However, when we look at the diagram as a whole we can see a different drop in the Li content 
with magnitude for the two populations. Specifically, while Li seems to have a gentle
decrease with luminosity for the Na-poor stars, the Na-rich group presents a very abrupt decline, i.e., at the bump luminosity Li suddenly
disappears. This fact, which reflects different timescales for mixing and 
hence for Li depletion\footnote{We estimated the $e$-folding time for Li abundance using the tracks by Bertelli et al. (2008).
Na-poor stars reduce to about a factor of 2 their Li content in 0.17 mag , which corresponds to $\approx$ 10 Myr
(this is indeed the time required by a 0.8 $M_\odot$ star to
become brighter by 0.17 magnitudes after it has left the RGB bump). For the Na-poor stars, this time is smaller by at least an order of
magnitude, i.e. $\leq$2Myr.}, suggests
a structural difference between Na-rich and Na-poor stars; however, no current theoretical model predicts such a
behavior and we cannot provide a satisfactory explanation to date. In this context, we mention that 
 the so-called 
``thermohaline mixing"  has been proposed responsible for non-canonical mixing acting at the RGB bump
(see, e.g., Eggleton et al. 2007; Charbonnel \& Zahn 2007). Eggleton and coworkers suggested that the molecular weight
inversion created by the $^3$He($^3$He,2$p$)$^4$He reaction could be the cause of such a mixing: why 
Na-rich and Na-poor stars should be differentially affected by this kind of mixing is not obvious
and our result could be the input for further theoretical and observational investigations in this
direction.

\begin{center}
\begin{figure}
\includegraphics[width=9cm]{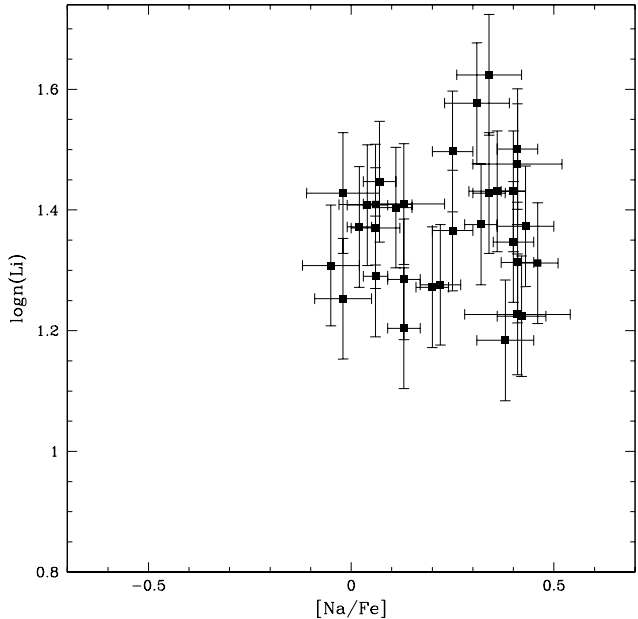}
\caption{Li vs. Na for stars below the luminosity bump. Error bars for Na come from
Marino et al. (2008); the uncertainties in Li are due to errors on 
best fit and effective temperatures.}\label{f:lina}
\end{figure}
\end{center}

\begin{center}
\begin{figure}
\includegraphics[width=9cm]{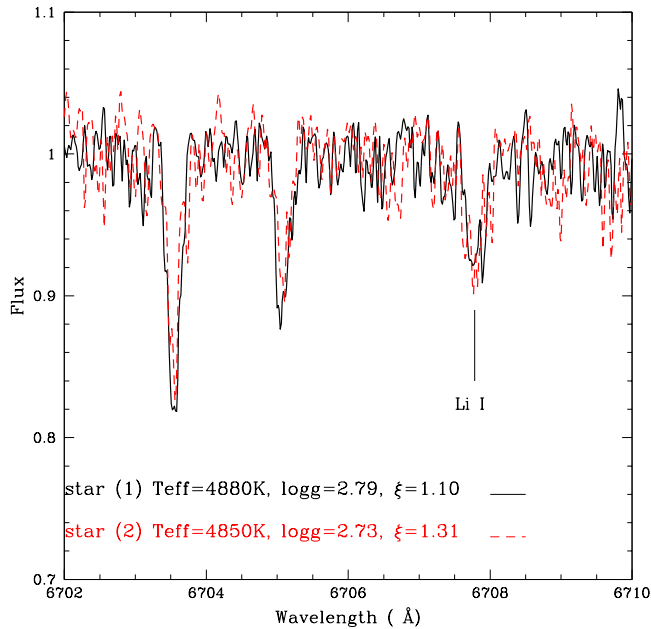}
\caption{Comparison of two spectra, near the Li doublet at 6708 \AA, for the two
extreme cases for Na abundances. Solid and dashed lines are for [Na/Fe]=-0.02
and [Na/Fe]=+0.43, respectively. (A color version of this figure is available in the online journal.)}\label{f:spec}
\end{figure}
\end{center}

By considering only the stars fainter than the bump luminosity, we show Li abundances as 
a function of Na in Figure~\ref{f:lina}: as one can see, there is no Li$-$Na anti-correlation, with second generation stars
([Na/Fe]$>$0.2 dex) sharing the same Li content of the primordial population.
As an example, we show in Figure~\ref{f:spec} the spectra, around the Li~{\sc i} region, 
for the two most extreme cases in Na abundances. The two stars, with [Na/Fe]=$-$0.02 and 
[Na/Fe]=+0.43, respectively, show identical Li features (note that the stars have very similar
parameters, and the same line strength reflects the same Li abundance). 
The average Li abundances are $\log{n{\rm (Li)}}$=1.336$\pm$0.023 (rms 0.062)
and $\log{n{\rm (Li)}}$=1.387$\pm$0.038 (rms 0.136), respectively for
Na-poor and Na-rich stars: although we derive the same Li abundance for the two populations, 
it is interesting to note that Na-rich stars have a larger scatter in Li with respect to Na-poor ones. 
A one-tailed Fisher test returns a 5\% probability such that a difference can be 
obtained by chance.

{\it A natural explanation for this similarity in Li content between first and second generation stars (with the last
showing a larger scatter) is a Li production.  } 
In fact, if a decrease in O of $\sim$50\%/60\% occurred (as derived e.g. by Marino et al.
2008 and Carretta et al. 2010), also Li must have been depleted.
In a recent work, D'Antona \& Ventura (2010) have presented the expected Li production as a function of the polluter mass
(AGB stars) for metallicity $Z$=0.001. 
Looking at their Figure 5, one can see that a very low mass AGB polluter
(i.e., $\approx$4~$M_\odot$) can produce a moderate Li content with values very close 
to the Li plateau ($\log{n{\rm (Li)}}$$\sim$2.2-2.3). 
After considering a depletion of a factor of $\sim$20 at the 1DUP, this result
agrees very well with our values (i.e. $\log{n{\rm (Li)}}$$\sim$1.3-1.4).
As also briefly explained in the Section~\ref{sec:intro} (and widely discussed in Carretta et al. 2010), there are 
further indications
that only low-mass polluters could have contributed to the observed chemical pattern in M4: (1) an almost
``vertical" Na$-$O anticorrelation, with very small oxygen variation (depletion);
and (2) the lack of Mg$-$Al anticorrelation\footnote{
We note that Marino et al.\ (2008) found evidence for a small increase ($\sim$0.10 dex) of Al 
with Na, with Na-rich stars also slightly Mg depleted (see their Table 7). 
Evidence for a Na$-$Al correlation was also found by Ivans and coworkers. In any case, 
the Al variation is small, and no 
Mg$-$Al anticorrelation has been observed among M4 stars by Ivans et al.\ (1999), 
Marino et al.\ (2008), and Carretta et al.\ (2009b). Also the Mg isotope ratios, as derived by Yong et al. (2008), show
no variation within M4 with values very close to solar ones: the same authors concluded that this is not surprising
due to the very little Al variation in this GC.},
which in fact requires high mass polluters for the activation of higher temperature cycles
($T\sim$65~MK; Prantzos \& Charbonnel 2006).
A similar case could have occurred for NGC~6397, where according to Pasquini et al. (2008), two stars 
differ by $\sim$0.6 dex in O, but have the same ``normal" Li ($\log{n{\rm (Li)}}$=2.2). 
 Also, in a recent work Lind et al. (2009), based on a sample of $\sim$100 MS and early SGB/stars, found
no difference in Li abundances between Na-rich and Na-poor stars with only two stars driving a Li$-$Na anticorrelation. 
They concluded that Li content is independent of intra-cluster pollution; however the Na$-$O distribution
points out to a certain (though small) degree of oxygen depletion and, as a consequence, of Li destruction as well.
Hence, if first and second generation stars share the same Li abundances, a Li production should also be required 
for this cluster.
Along with 
a difference in metallicity of $\sim$0.9 dex, the two
clusters, NGC~6397 and M~4, have both quite small integrated magnitudes (i.e., mass) 
with $M_{\rm Vt}$=$-$6.63 and $M_{\rm Vt}$=$-$7.20, respectively (Harris 1996). 
The similarity in masses between these two GCs also seems to suggest a similar ``typical" polluter for both M4 and 
NGC~6397, with the requirement to have in both cases neither very high mass polluters (no extended MgAl/NaO anticorrelations,
very little He enhancement\footnote{Although He is the main product of hot H-burning through CNO cycle, the relationship between
He and $p$-capture elements is quite complex. For ``extreme" GCs, like NGC~2808 and $\omega$Cen (with
extended NaO-MgAl anticorrelations, multiple MSs and peculiar HB morphology) the $Y$ content can reach up to $\approx$0.40.
For all the other GCs, even if they show the NaO anticorrelations, large differences in $Y$ are not necessary
(see for details Gratton et al. 2010).}, and no Li$-$Na anticorrelation) nor low mass polluters ($\leq$4 $M_\odot$),
otherwise the C+N+O is not constant and/or $s$-process variations should be present (see Ivans 1999; Yong et al. 2008).

The more massive GC NGC~6752 ($M_{\rm Vt}$=$-$7.73) could present a different behavior.
We might speculate that 
only a very low Li production from higher mass polluters of $\approx$5$-$6$M_\odot$ (see Figure 5 of
D'Antona \& Ventura, 2010) does not erase Li$-$Na anticorrelation for this cluster\footnote{The trend of Li production with polluter 
mass is not linear
(see Figure~5 of D'Antona \& Ventura, 2010): the low mass polluters (3.5$-$4.5 $M_\odot$ 
show a Li production close to the Plateau values,
while in 
the higher masses regime (5$-$7 $M_\odot$) the Li yields become smaller (a ``v-like" distribution). 
Moving to very high-mass stars
(7-8 M$_\odot$), the Li production reaches extremely high values, up to $\log{n{\rm (Li)}}$ $\approx$4 dex.}.
Note in fact that NGC~6752 presents an extended Na$-$O anticorrelation and 
a large variation in Al (i.e., the MgAl chain was active in the polluter stars).

On the other hand, it seems very difficult to discriminate the nature of polluters and their properties
for 47 Tuc: maybe this GC was similar to NGC~6752 but the intrinsic scatter in Li abundance, independent 
of intracluster pollution, washes out the fossil imprint by the previous generation of polluter stars (D'Orazi et al. 2010).
Given the large uncertainties linked to model predictions (cross sections, mass
loss law, overshooting, convection treatment, etc.), 
here our general aim is to provide a ``qualitative" comparison between theoretical prescriptions
by D'Antona \& Ventura (2009) and observational evidence. Also, we stress that our result, based on only a few objects, needs to be confirmed by 
including a larger number of clusters and of star per
cluster. However, we think that our data provide a quite robust 
observational evidence of AGB stars as responsible for GC pollution.

\section{SUMMARY AND CONCLUSIONS}\label{sec:sum}  
We report in this Letter Li abundances for a sample of $\sim$100 giants, both below and above the RGB bump luminosity, 
belonging to the GC NGC~6121 (M4). The main purpose of our work was the study of the correlation (if any) 
between Li content and elements involved in $p$-capture reactions, the last ones already presented for the very same stars by Marino et al.
(2008). The principal results we obtained in our investigations can be summarized as follows.

\begin{itemize}
\item[1.] As expected, Li tends to disappear as the stars reach the RGB bump luminosity; however the Na-rich and Na-poor stars 
show very different trend of Li with magnitude. Specifically, while the decline of Na-poor stars 
with $M_{\rm V}$ is rather 
smooth, there is a very brusque decrease for Na-rich stars. The so-called ``thermohaline mixing", which seems responsible for 
extramixing processes at the RGB bump, is not predicted to have different outcomes in Na-poor and Na-rich stars. 
Further investigations of this aspect are mandatory both from observational and theoretical points of view. New observations, 
focusing on Li determination along the RGB in several GCs, are necessary to assess if M4 is a ``peculiar" case or  
other GCs share the same behavior; as a consequence model predictions could be revised in this sense.
\item[2.] M4 does not show any Li$-$Na anticorrelation, with first and second generation stars having
almost the same Li content. Along with similarities in Li abundances, the larger scatter found in Na-rich stars, indicates that a Li
production, from the previous generation of polluters, must have happened. This provides support to intermediate-mass AGB stars
responsible for intra-cluster contamination, since FRMS can only destroy Li.

\end{itemize} 

\begin{table*}
\caption{Stellar Parameters and Abundances for Our Sample Stars. This table is available in its entirety in a machine-readable form in the
online journal. A portion is shown here for guidance regarding its form and content.}\label{t:tab1}
\begin{center}
\begin{tabular}{lccccccc}
\hline\hline
star  & T$_{\rm eff}$ & log$g$ & $\xi$       & [Na/Fe] & err$_{\rm Na}$ & log~n(Li) & err$_{\rm Li}$ \\
      &   (K)         &        &(kms$^{-1}$) &         &                &           &               \\
\hline
30345 & 4850          & 2.73   &1.31         &  0.43   & 0.07 	        &  1.37     & 0.08\\
30452 & 4830          & 2.56   &1.25         &  0.06   & 0.03 	        &  1.29     & 0.09\\
30719 & 4810          & 2.65   &1.24         &  0.42   & 0.06 	        &  1.22     & 0.08\\
31306 & 4900          & 2.87   &1.33         &  0.40   & 0.05 	        &  1.35     & 0.07\\
 \hline\hline
\end{tabular}
\end{center}
\end{table*}

\acknowledgments
We warmly thank Raffaele Gratton, Angela Bragaglia, 
Eugenio Carretta and Sara Lucatello for very helpful 
suggestions and useful comments in all the stages of the preparation of this Letter.
This work was funded by
the Italian MIUR under PRIN 20075TP5K9. The anonymous referee is kindly
acknowledged for very valuable comments which clarified the manuscript.

{}
\end{document}